\documentclass{scspaperproc}

\usepackage{latexsym}
\usepackage{graphicx}
\usepackage{mathptmx}

\usepackage{amsmath}
\usepackage{amsfonts}
\usepackage{amssymb}
\usepackage{amsbsy}
\usepackage{amsthm}

\usepackage{algorithm}
\usepackage{algpseudocode}
\usepackage{pifont}
\usepackage{graphicx}
\usepackage{float}
\usepackage{times}
\usepackage{amsmath}
\usepackage{varwidth}

\usepackage[pdftex,colorlinks=true,urlcolor=blue,citecolor=black,anchorcolor=black,linkcolor=black]{hyperref}

\usepackage{hyphenat}
\newtheoremstyle{scsthe}
{8pt}
{8pt}
{\it}
{}
{\bf}
{.}
{.5em}
{}

\theoremstyle{scsthe}

\begin{document}

%
%

\pagestyle{fancyplain}

\thispagestyle{plain}
\firstPageHead{}

\chead{\fancyplain{}{\itshape Datta, Tallamraju, and Karlapalem \vspace{8pt}}}

\rhead{}
\cfoot{}
\renewcommand{\headrulewidth}{0pt} 

\makeatletter
\let\@internalcite\cite
\def\cite{\def\@citeseppen{-1000}%
    \def\@cite##1##2{(##1\if@tempswa , ##2\fi)}%
    \def\citeauthoryear##1##2##3{##1 ##3}\@internalcite}
\def\citeNP{\def\@citeseppen{-1000}%
    \def\@cite##1##2{##1\if@tempswa , ##2\fi}%
    \def\citeauthoryear##1##2##3{##1 ##3}\@internalcite}
\def\citeN{\def\@citeseppen{-1000}%
    \def\@cite##1##2{##1\if@tempswa, ##2)\else{}\fi}%
    \def\citeauthoryear##1##2##3{##1 (##3)}\@citedata}
\def\citeA{\def\@citeseppen{-1000}%
    \def\@cite##1##2{(##1\if@tempswa , ##2\fi)}%
    \def\citeauthoryear##1##2##3{##1}\@internalcite}
\def\citeANP{\def\@citeseppen{-1000}%
    \def\@cite##1##2{##1\if@tempswa , ##2\fi}%
    \def\citeauthoryear##1##2##3{##1}\@internalcite}
\def\shortcite{\def\@citeseppen{-1000}%
    \def\@cite##1##2{(##1\if@tempswa , ##2\fi)}%
    \def\citeauthoryear##1##2##3{##2 ##3}\@internalcite}
\def\shortciteNP{\def\@citeseppen{-1000}%
    \def\@cite##1##2{##1\if@tempswa , ##2\fi}%
    \def\citeauthoryear##1##2##3{##2 ##3}\@internalcite}
\def\shortciteN{\def\@citeseppen{-1000}%
    \def\@cite##1##2{##1\if@tempswa, ##2\else{}\fi}%
    \def\citeauthoryear##1##2##3{##2 (##3)}\@citedata}
\def\shortciteA{\def\@citeseppen{-1000}%
    \def\@cite##1##2{(##1\if@tempswa , ##2\fi)}%
    \def\citeauthoryear##1##2##3{##2}\@internalcite}
\def\shortciteANP{\def\@citeseppen{-1000}%
    \def\@cite##1##2{##1\if@tempswa , ##2\fi}%
    \def\citeauthoryear##1##2##3{##2}\@internalcite}
\def\citeyear{\def\@citeseppen{-1000}%
    \def\@cite##1##2{(##1\if@tempswa , ##2\fi)}%
    \def\citeauthoryear##1##2##3{##3}\@citedata}
\def\citeyearNP{\def\@citeseppen{-1000}%
    \def\@cite##1##2{##1\if@tempswa , ##2\fi}%
    \def\citeauthoryear##1##2##3{##3}\@citedata}
%
%
%
\def\@citedata{%
    \@ifnextchar [{\@tempswatrue\@citedatax}%
                  {\@tempswafalse\@citedatax[]}%
}

\def\@citedatax[#1]#2{%
\if@filesw\immediate\write\@auxout{\string\citation{#2}}\fi%
  \def\@citea{}\@cite{\@for\@citeb:=#2\do%
    {\@citea\def\@citea{, }\@ifundefined
       {b@\@citeb}{{\bf ?}%
       \@warning{Citation `\@citeb' on page \thepage \space undefined}}%
{\csname b@\@citeb\endcsname}}}{#1}}%

%
\def\@citex[#1]#2{%
\if@filesw\immediate\write\@auxout{\string\citation{#2}}\fi%
  \def\@citea{}\@cite{\@for\@citeb:=#2\do%
    {\@citea\def\@citea{, }\@ifundefined
       {b@\@citeb}{{\bf ?}%
       \@warning{Citation `\@citeb' on page \thepage \space undefined}}%
{\csname b@\@citeb\endcsname}}}{#1}}%

%
\def\@biblabel#1{}
\makeatother

\newdimen\bibindent
\bibindent=.25in

\def\thebibliography#1{\section*{\refname}\list
   {}{\settowidth\labelwidth{[#1]}
   \leftmargin \bibindent
   \itemindent -\bibindent
   \listparindent \itemindent
	 \itemsep 4pt
   \parsep 0pt
   \usecounter{enumi}}
   \def\newblock{}
   \sloppy
   \sfcode`\.=1000\relax}

\setlength{\baselineskip}{12.7pt}

\def\SCSconferenceacro{SummerSim}

\def\SCSpublicationyear{2019}

\def\SCSconferencedates{July 22-July 24}

\def\SCSconferencevenue{Berlin, Germany}

\def\SCSsymposiumacro{SCSC} 

\title{Multiple Drones driven Hexagonally Partitioned Area Exploration: Simulation and Evaluation}


\author{
Ayush Datta \\
Rahul Tallamraju \\
Kamalakar Karlapalem\\ [12pt]
Agents and Applied Robotics Group\\
International Institute of Information Technology, Hyderabad \\
Gachibowli, Hyderabad, India \\
\{ayush.datta,rahul.t\}@research.iiit.ac.in, kamal@iiit.ac.in\\
}

\maketitle

\section*{Abstract}

In this paper, we simulated a distributed, cooperative path planning technique for multiple drones ($\sim$200) to explore an unknown region ($\sim$10,000 connected units) in the presence of obstacles. The map of an unknown region is dynamically created based on the information obtained from sensors and other drones. The unknown area is considered a connected region made up of hexagonal unit cells. These cells are grouped to form larger cells called sub-areas. We use long range and short range communication. The short-range communication within drones in smaller proximity helps avoid re-exploration of cells already explored by companion drones located in the same subarea. The long-range communication helps drones identify next subarea to be targeted based on weighted RNN (Reverse nearest neighbor). Simulation results show that weighted RNN in a hexagonal representation makes exploration more efficient, scalable and resilient to communication failures.

\textbf{Keywords:} Multi Robot Coordination, Drones, Cooperation

\section{Introduction}
\label{sec:intro}

\subsection{Motivation}

These days, drones have been used increasingly for exploration of an unknown area.  There are various applications of area exploration like search and rescue, map building, intrusion detection, and planetary exploration. Building co-ordinated multiple basic drones is cheaper than one expensive and complex drone. The idea behind using co-ordinated drones is that if one drone takes a certain amount of time to explore a region, then two co-ordinated drones should take half the time taken by a single drone. Similarly, if more drones are exploring a certain region, they should reduce the exploration time drastically. Assuming that some identical drones are equipped with sensing, localization, mapping and communication capabilities, they need to explore the unknown area efficiently and reliably. \par
The challenge of exploring an unknown area with obstacles is that the drones tend to explore the same area multiple times, which further tends to increase due to lesser communication between them. Since centralized coordination algorithms usually suffer from a single point of failure problem, it is desirable to have a distributed co-ordinated system to improve system reliability. \par
The drones have their own memory in which they keep track of the area explored and the information about the position of the obstacles. If they do not communicate this information among themselves, then they tend to explore the area which is already covered by the other drone. This further increases the exploration time. 
If they communicate less and exchange less information, then it leads to a more redundant exploration of the cells. At the same time, we need to reduce the amount of communication between the drones and make the system more resilient to communication failures. Further, the mechanism of co-ordination should be such that it is scalable to a large area with multiple drones. \par
Efficient exploration of the area means that drones do not explore the same area multiple times and they explore the complete area in a minimum amount of time and with minimum drone traveling distance and with minimal communication between them.
To address these issues, we propose Hexagonally Partitioned Area Exploration using Reverse Nearest Neighbors. It is a modification of the distributed frontier based algorithm ~\cite{wang2011frontier} with Particle Swarm Optimization (PSO). In this paper, we show that our solution outperforms the PSO model based approach. \par
The area to be explored is tessellated into hexagonal subareas, unlike the traditional quadrangular grid. First, the drones explore their own subarea using the traditional frontier based algorithm with a limited or short range communication with drones present in the same subarea.
After exploration of the own subarea, the drones decide which subarea to explore based on weighed reverse nearest neighbor (RNN). This is aided by the long range communication to figure out which drones are planning to approach which subarea. The RNN assigns the drones to the subareas in such a way that the communication and area exploration time reduces drastically. 

\subsection{Previous Work}	
The initial work in the field of area exploration started with a single autonomous robot with sensing, localization and mapping. The work on multi-robot exploration was started by \citeN{yamauchi1997frontier} who introduced distributed frontier algorithm. Since there is no explicit co-ordination among the robots, they tend to move towards the same frontier cells which introduces inefficiency. Based on similar frontier concept, \shortciteN {simmons2000coordination} developed a semi-distributed multi-robot exploration algorithm which requires a central agent to evaluate the bidding from all the other robots to obtain the most information gain while reducing the cost, or the traveling distance.  \shortciteN{berhault2003robot} used Combinatorial Auction for the area exploration. The disadvantage of the bidding algorithms is that they have a single point of failure and since the communication cost increases drastically with the increase in the number of robots, these algorithms are not scalable.\par
There has also been work on heterogeneous robots exploring the area by \citeN {singh1993map}. However, they do not focus on efficiency. Market based map exploration was also proposed by \shortciteN{zlot2002multi}. Ant-inspired algorithm to divide the area into square cells on which the robots leave trails of their passage was done by \citeN{koenig2001terrain}. All these strategies require a central shared memory where the robots can leave their marks in the environment. This introduces a single point of failure which makes the system less robust. \citeN{fu2009local} made use of local Voronoi decomposition for task allocation in map exploration. \par
\citeN{kennedy1995particle}, proposed the PSO model which was inspired by the earlier research by bird flocks. \citeN{al2008optimizing} added the power cost as the indicator in PSO model to the WSN optimization. \citeN{dasgupta2009flocking} built a robot team based on Reynolds' flocking model to improve the efficiency of the multi-robot map exploration. \citeN{wang2011frontier} proposed PSO model for frontier based exploration. \citeN{dornhege2013frontier} combined the concept of voids with frontier based approach to search for entombed victims in confined structures. \citeN{yoder2016autonomous} improvises frontier algorithm performance by using only state-changed space in the 3D map in each iteration. \citeN{mahdoui2018cooperative} modifies frontier based approach where instead of sharing local maps, robots share their local frontier points with a so-called Leader.\par

In \textbf{Quadrangular RNN}, we show that RNN outperforms the traditional PSO model as it keeps a track of which drones are approaching which subarea instead of keeping track of drones that are in the path of the subarea to be explored. As proposed by \shortciteN{sheng2006distributed}, we need to use a nearness measure so that the agents remain within the communication radius. This approach leads to duplicate exploration of the area. To overcome this, in \textbf{Quadrangular RNN with SRC} we show that using two kinds of communication improves the exploration efficiency. Recently, a lot of research has been done on Multi-Robot area exploration and swarm intelligence, but most of these algorithms use a quadrangular grid to divide the area into subareas. In \textbf{Hexagonal RNN  with SRC}, we show that using a hexagonal grid over a quadrangular grid reduces the exploration time as there are six degrees of directions to move instead of four. Finally, \textbf{Weighted HRNN with SRC} also highlights how this system is failure resilient, scalable and avoids the possible deadlocks.

\section{Problem Formulation}
We assume that identical drones are equipped with sensing, localization and communication capabilities. The task is to explore an unknown area avoiding the obstacles efficiently and reliably, at the same time making it resilient to communication failure and hardware malfunctioning of the drone. Each drone is assumed to have its own memory which it updates after every step of exploration. In our testing scenario, the obstacles are randomly distributed in the bounded exploring area and each drone is assumed to correctly locate itself on the map. We have divided the complete environment into either square cells or hexagonal cells. Moreover, each drone is equipped to communicate and share information with companion drones. \par

In case of the map being divided into a quadrangular grid, we assume that the drones are equipped with eight sensors that can sense the environment in eight directions. The sensors are equidistant around the drone, so the drone can detect local environment in eight directions: Front, Right-front, Right, Right-back, Back, Left-Back, Left, and Left-front. The drones can move only in four directions: Front, Right, Back, Left.
In case of the map being divided into hexagonal cells, we assume that the drones are equipped with six sensors which can sense the environment. The hexagonal grid allows the drones to move in six directions around it. \par
Each drone stores the environment in its memory in a cell based map. The cell based map is stored in a \textit{[n x m]} matrix for both hexagonal and square grid. Each cell in the map is either \textbf{unexplored, obstacle, visited or it is a frontier cell}. The frontier cell is the cell which has been sensed by drone, but it has not yet been visited. The whole environment is divided into subareas either in rectangular regions composed of square cells in it or in hexagonal subareas composed of hexagonal cells in it. \par
The drones are initially placed randomly into cells. The goal is to explore all the cells in the area at least once. The drones can communicate with companion drones and co-ordinate to reduce the time to cover the complete area. The communication allows the robots to exchange their local maps with the companion drones.

\section{Multi-Robot Area Exploration Algorithm}
We present the algorithm in four steps where at each step we modify the algorithm in such a way such that the efficiency is progressively improved. \par
In all the approaches we divide our algorithm into two stages. Initially, the drones are randomly distributed among the subareas. In the first part, \textbf{Exploration state} the drones explore their own local subareas and later in the second part, \textbf{Moving State} the drones move towards the unexplored cells in the different subarea determined by RNN or PSO model. Each drone runs its own algorithm and maintains exploration or moving state. It is done in such a way that the total time to explore the complete area is minimized. 

\subsection{\textbf{Quadrangular Reverse Nearest Neigbour (QRNN)}}
\subsubsection{\textit{Exploring local subarea (Exploration state)}}
In the exploration state, the drones keep moving towards the frontier cells recording the terrain of the nearby cells until there are no frontier cells left in the same subarea. \par
Initially, the cell where a drone is located is marked visited and the neighboring cells are marked frontier cells or obstacle cells in the local memory of each drone. After each time step, the drones move to the nearby frontier cell marking it covered and marking the new neighbor cells as frontier cells. The drone selects the frontier cell to move in this order {North, South, East and, West}. If there are no more neighboring frontier cells, then the drones move towards the nearest frontier cell. This goes on until there are no more frontier cells left in the subarea. The complete details are written in Algorithm1.\par
To find the nearest frontier cell in a grid with obstacles in between, we use A* search to find the nearest frontier cell.\par
The function of the A* search is given by:
\begin{equation}\label{eq:star}
f(i,j) = g(i,j) + h(i,j)
\end{equation}

In equation \eqref{eq:star}, $g(i,j)$ is the distance along the shortest path that connects the two cells \textit{`i'} and \textit{`j'}. Here $h(i,j)$ is the displacement or crow fly distance between the two cells as explained by \citeN {nosrati2012investigation}. \autoref{Subarea Coverage} explains the Exploration State.

\begin{algorithm}
\caption{Subarea Coverage}
\label{Subarea Coverage}
\begin{algorithmic}[1]
\Procedure{Sub\textendash Area Coverage}{}
\State Mark the current cell as a visited cell.
\State 
	Mark the neighbor unexplored cells in the same subarea as frontier cells or obstacles.
\If{there are frontier cells around}
	\State go to one of them 
\ElsIf{frontier cells in the same subarea present}
	\State go to nearest frontier cell along the shortest path
\Else{ Moving state}
\EndIf
\State go to 2

\EndProcedure
\end{algorithmic}
\end{algorithm}
\subsubsection{\textit{RNN based co-ordination (Moving State)}}
In this state, the drones have to move towards an appropriate subarea. The drones should move towards a subarea ensuring that a minimal number of companion drones are approaching the same subarea.\par
Previous papers use PSO based cooperation model to avoid multiple drones heading towards the same cell. The PSO model uses the velocity which is updated according to previous best positions and the global best position achieved by its neighbors. The global best position is defined as the cell with a minimum number of other drones in its direction.
\par
\begin{figure}[htb]
{
\centering
\includegraphics[width=0.50\textwidth]{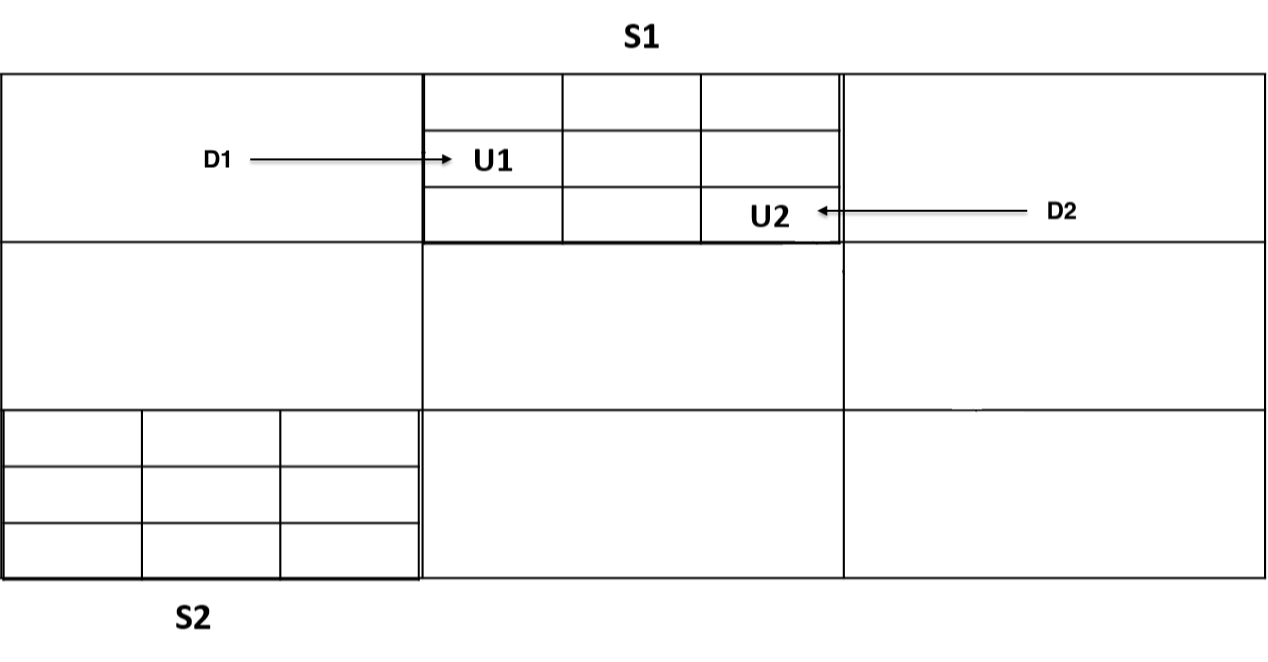}
\caption{Quadrangular Grid of size 9X9 with 9 subareas.}\label{fig:Picture1}
}
\end{figure}

 
The disadvantage of the PSO model and can be explained using \autoref{fig:Picture1}.

In the above quadrangular grid, $D_{1}$ and $D_{2}$ represent the drones, $S_{1}$ and $S_{2}$ are unexplored subareas and $U_{1}$, $U_{2}$ are the unit cells of subarea  $S_{1}$. Considering a case where only subareas $S_{1}$ and $S_{2}$ have unexplored cells left and rest all area is explored.
If we follow the PSO model, then both drones $D_{1}$ and $D_{2}$ will move towards $S_{1}$, as unexplored cells $U_{1}$, $U_{2}$ are more closer to them than the cells of subarea $S_{2}$ but ideally, since there is nobody to move to unexplored region $S_{2}$, the $D_{1}$ should move towards $S_{2}$ while the drone $D_{2}$ should move towards $S_{1}$.
\par
To overcome this problem, we use the concept of reverse nearest neighbor. The intuition behind this concept is to keep a track of the number of drones approaching that subarea. If a greater number of drones are approaching that subarea then the cost to travel to that subarea should be increased. RNN outperforms the traditional PSO model as it keeps a track of which drones are approaching which subarea instead of keeping track of drones that are approaching a particular cell. We call it reverse nearest neighbor because instead of focusing on the nearest area to a drone we focus on the nearest drone to a subarea.

The cost for the Drone $D_{i}$  at $i$ to travel to a cell $U_{j}$ at $j$ is calculated as shown in \eqref{eq:star1}:
\begin{equation}\label{eq:star1}
Cost (D_{i} , U_{j} ) = g(i , j) + h(i , j) + \alpha*S_{(U_{j})}
\end{equation}
Here, $S_{(U_{j})}$ is the number of drones that were approaching the subarea in which $U_{j}$ is located. $\alpha$ is the constant just to scale this factor. Functions \textit{`g'} and \textit{`h'} are part of A* search as explained earlier.
\par

Using RNN, we can see that initially, both the drones will move towards the nearby unexplored region $S_{1}$ but after a time step, the cost in RNN for $D_{1}$ to move towards $S_{1}$ will increase thereby, pushing $D_{1}$ towards $S_{2}$ and $D_{2}$ towards $S_{1}$.

Long range communication facilitates that all the drones exchange information towards which subarea they tend to approach after each time stamp. Even if the communication fails with some far away drones, then also the algorithm keeps working based on the previous information. Hence, this algorithm is distributed and resilient to communication failure.

\begin{algorithm}[htb]
\caption{RNN based coordination}
\label{RNN based coordination}
\begin{algorithmic}[1]
\Procedure{RNN based coordination}{}
\State Initialise, drones are not approaching any subarea in present iteration.
\State Communicate(Long Range) with companion drones to find which subarea they were approaching \newline
\hspace*{1.25em} in previous iteration.
\State Calculate the cost $Cost (D_{i} , U_{j} )$ to approach each cell based on number of drones $S_{(U_{j})}$ moving\newline
\hspace*{1.25em} towards subarea of $U_{j}$ in previous iteration.
\State Move towards the cell with minimum cost.
\State Update which subarea drone is approaching.
\State go to Exploration State (Algorithm 1)

\EndProcedure
\end{algorithmic}
\end{algorithm}

In \autoref{RNN based coordination}, the drone in \textbf{Moving State} makes a step towards selected subarea.
After making a step it comes back to \textbf{Exploration State} if there are frontier cells around it.

\subsection{\textbf{QRNN with Short Range Communication (QRNN-SRC)}}
This approach is similar to the previous approach, showing and improving upon the demerits of QRNN. In QRNN during the exploration state, it is possible to have multiple drones in the same subarea. This is more likely if the number of drones exploring the area is high. In QRNN, communication was only taking place in the movement state. The drones might follow one another to explore their own subarea. As all the drones follow the order of direction to move that is {North, South, East and, West}, the drones might end up covering a lot of cells multiple times.

\begin{figure}[htb]
{
\centering
\includegraphics[width=0.50\textwidth, height=0.30\textwidth]{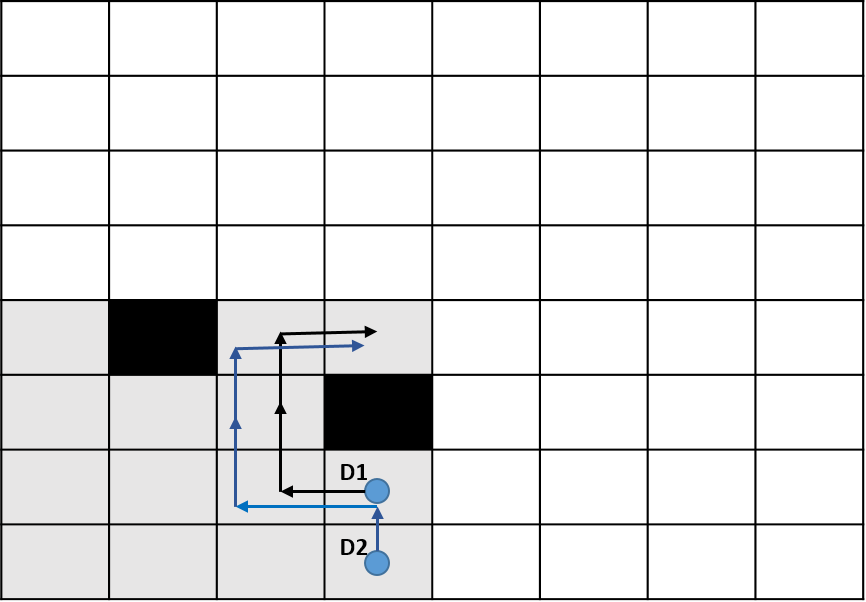}
\caption{Quadrangular Grid.}\label{fig:Picture2}
}
\end{figure}

As shown in \autoref{fig:Picture2} if the drones are in the same subarea, then one drone might follow the path of a companion drone as the other drone is unaware of what the first drone has already explored.

Therefore, drones within a subarea need to communicate more frequently. The short range communication amongst the drones in smaller proximity helps avoid obstacles and re-exploration of cells already explored by companion drones located in the same subarea.
The short range communication can also be multi-hop communication between the nearby drones leading to cascading information exchange. 
\par
The long range communication like in QRNN helps in deciding optimal next sub area to be targeted by individual drones based on RNN (Reverse nearest neighbor).

\subsection{\textbf{Hexagonal RNN with SRC (HRNN-SRC)}}
One of the most common and simplest representations of the area to be explored is the quadrangular grid map. In a quadrangular grid, the diagonal distance between two squares is bigger than the horizontal or vertical distance between two squares. Thus, it is harder to manage distances when drones move in diagonal, vertical and horizontal ways.
\par
To simplify that problem, drones are frequently restricted to only make vertical or horizontal moves when quadrangular map representation is used. In a hexagonal grid, diagonal and vertical distances between two hexagons are the same. Thus, there is no need to worry about managing different distances scales or restrict drones to only make vertical or horizontal moves. Therefore, in this approach, we use the hexagonal cells and hexagonal subareas to represent the area to be explored.
\par
The hexagonal approach gives drones six directions to move. Later simulation results show that there is an improvement in the efficiency using the hexagonal grid cells.

\subsection{\textbf{Weighted RNN-SRC (WHRNN-SRC)}}
In this approach, we improve upon the suggested RNN method in QRNN-SRC.
\par
\begin{figure}[htb]
\centering
\begin{minipage}{.5\textwidth}
  \centering
  \includegraphics[width=1.0\textwidth]{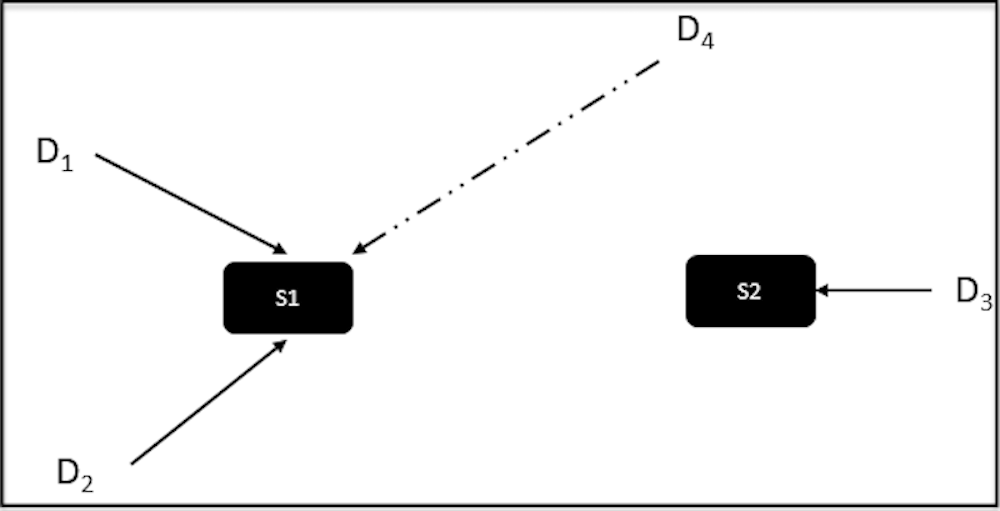}
  \caption{Drones are approaching subareas.}\label{fig:Picture3}
\end{minipage}%
\begin{minipage}{.5\textwidth}
  \centering
  \includegraphics[width=1.0\textwidth]{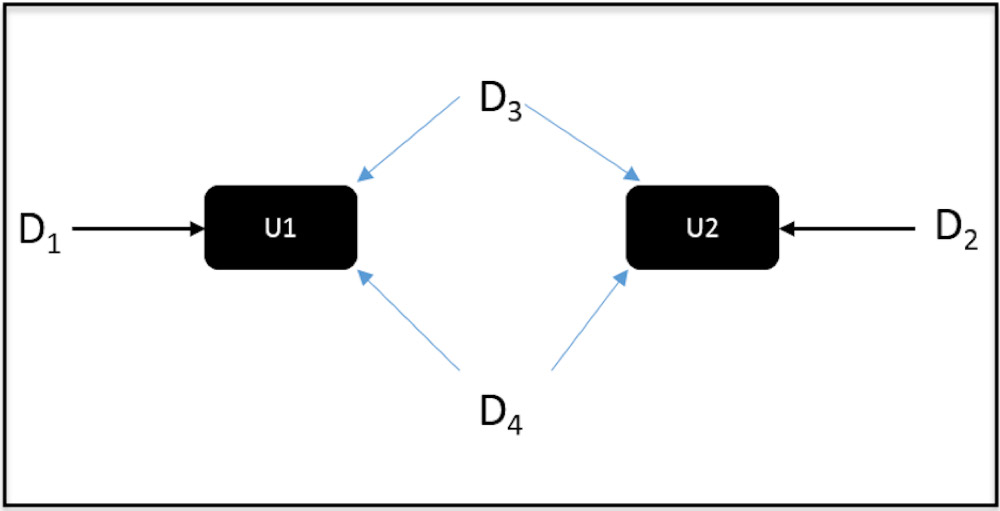}
  \caption{Deadlock Scenario.}\label{fig:Picture4}
\end{minipage}
\end{figure}

In the scenario, as shown in \autoref{fig:Picture3}, if there are drones $D_{1-4}$ and there are two subareas $S_{1}$ and $S_{2}$. More drones are approaching subarea 1, therefore using Reverse Nearest Neighbor $D_{4}$ will move towards $S_{2}$. But, if $S_{2}$ has only one cell left to be unexplored and the complete $S_{1}$ is unexplored then $D_{4}$ should move towards subarea 1. \autoref{fig:Picture4} explains the deadlock scenario which is explained later.

To address the above demerit, we propose a weighted RNN algorithm, wherein each subarea has a weight proportional to the number of cells presently unexplored in it. Now, the cost for the drone $D_{i}$ to travel to a cell $U_{j}$ is calculated as shown in formula \eqref{eq:star2}:
\begin{equation}\label{eq:star2}
Cost (D_{i} , U_{j} ) = g(i , j) + h(i , j) + \alpha*S(U_{j}) - \beta*\frac{W_{Sj}}{subarea.height*subarea.width}
\end{equation}
\par
Here, $W_{Sj}$ represents the number of cells unexplored in the subarea containing cell $U_{j}$ in the previous iteration.
The above formula penalizes agents more if they move towards a subarea where drones are already approaching by increasing cost. It also reduces the penalty if they move towards a completely unexplored subarea. $\beta$ is the constant which is multiplied by the percentage of unexplored cells in that subarea.
\par
The weighted RNN algorithm improves the efficiency keeping the system resilient to failure of communication. The efficiency is defined as minimizing number of times the cells are covered. 
\par
In this algorithm, we can see that towards the end of the simulation when only a few unexplored cells are left then, all the drones move towards few remaining unexplored cells unnecessarily. Hence, towards the end when the number of unexplored cells is comparable to the number of drones, we apply the bidding algorithm to further optimize the results.
\par
Since the RNN works at every time step, many times this can result in deadlocks.


In the scenario, as shown in \autoref{fig:Picture4}, initially the drones $D_{3}$ and $D_{4}$ move towards unexplored region $U_{1}$. In the next time step, since both of the drones proceed towards $U_{1}$. Now, using RNN the cost to the $U_{1}$ increases as two drones are trying to approach it. So, both the drones start moving towards unexplored area $U_{2}$. This way both the drones get stuck in the cycle causing a deadlock. To avoid this problem, we have used a timestamp mechanism, wherein the decision to move is delayed by an iteration. This way the deadlocks are also handled.
\section{Experimental Results}
We have tested our proposed approaches and compared them with the traditional PSO model. The experiments were simulated with a large number of grid cells (Max 10,000) and with a maximum of 200 drones. 
In the simulation experiment, we compared our approaches with the traditional PSO based model. 
\par
The various parameters for the simulation experiment are shown in \autoref{tab:first}:

\begin{table}[htb]
\centering
\caption{Parameters of the Simulation Experiment.}\label{tab:first}
\begin{tabular}{r|l}
\textbf{Parameters} & \textbf{Values}\\ \hline
Grid-size\{height, width\} & 40x40, 60x60, 80x80, 100x100\\ \hline
Size of Sub-area taken & \{4x4\},\{5x5\},\{8x8\},\{10x10\} \\ \hline
Number of Agents &  5, 10, 25, 50, 100, 150, 200\\  \hline
Number of Obstacles & 20\% \\\hline
Radius of short range communication (Euclidean Distance) & 5, 10, 15 , 20\\ \hline
Constant `$\alpha$' which gives weight to RNN & 2-10\\ \hline
\hline\end{tabular}
\end{table}

\begin{figure}[!htb]
\centering
\begin{minipage}{.5\textwidth}
  \centering
  \includegraphics[width=1.0\textwidth]{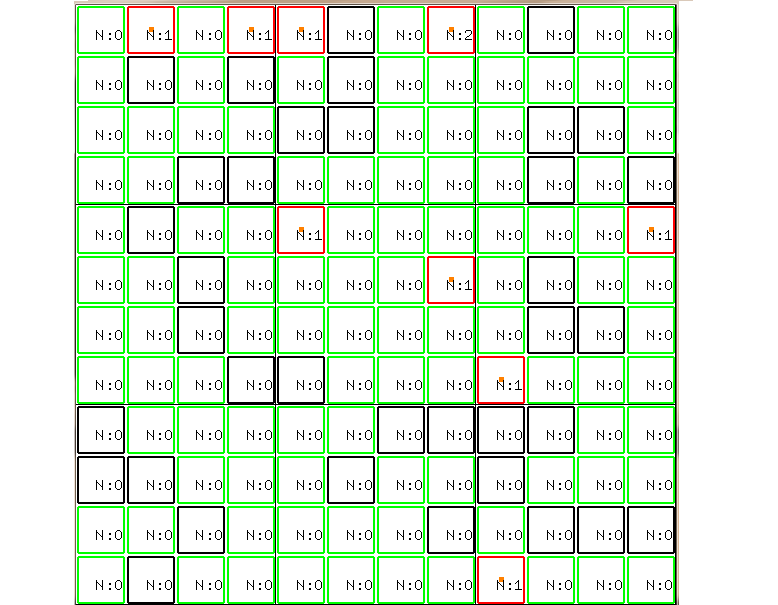}
  \caption{Quadrangular grid.}\label{fig:Picture5}
\end{minipage}%
\begin{minipage}{.5\textwidth}
  \centering
  \includegraphics[width=1.0\textwidth]{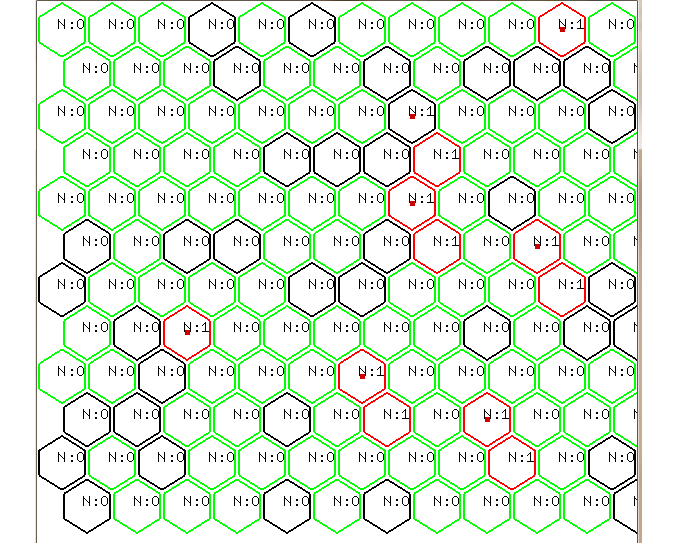}
  \caption{Hexagonal Grid.}\label{fig:Picture6}
\end{minipage}
\end{figure}

The traditional PSO based model, QRNN and QRNN-SRC are simulated on the quadrangular grid as shown in \autoref{fig:Picture5}. The HRNN-SRC and WHRNN-SRC are simulated on the hexagonal grid as shown in \autoref{fig:Picture6}.

\par
In \autoref{fig:Picture5} and \autoref{fig:Picture6}, the cells outlined in black color represents the obstacles, visited cells are outlined in red color and the unvisited cells are outlined in green color. The simulation environment shown above was created in SWIFT using OpenGL for visualization. The number inside the cell represents how many times that cell has been explored.
\par
\begin{figure}[htb]
{
\centering
\includegraphics[width=0.90\textwidth,height=0.39\textwidth]{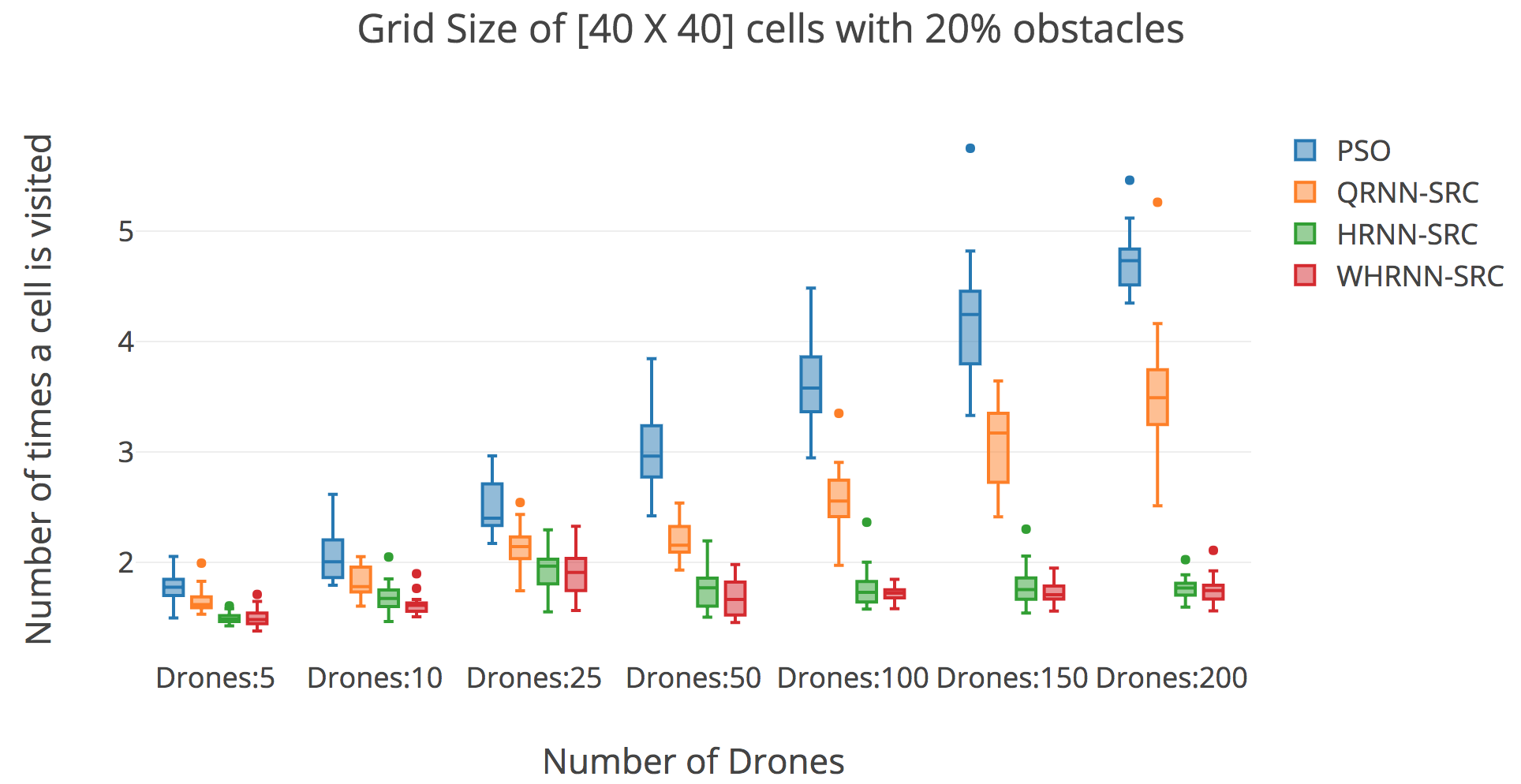}
\caption{Number of drones vs Average Re-explored Cells.}\label{fig:Picture7}
}
\end{figure}

\autoref{fig:Picture7} compares the performance of PSO, QRNN-SRC, HRNN-SRC and, WHRNN-SRC, with the increase in drones keeping the area to be explored constant. Here the grid size is 1600(40 x 40) cells.
It is clear from the figure that, for Weighted RNN, with increasing drones, the average number of cells re-explored is much less compared to the traditional PSO model. This proves that the algorithm is efficient and scalable. 

\begin{figure}[htb]
{
\centering
\includegraphics[width=0.90\textwidth,height=0.39\textwidth]{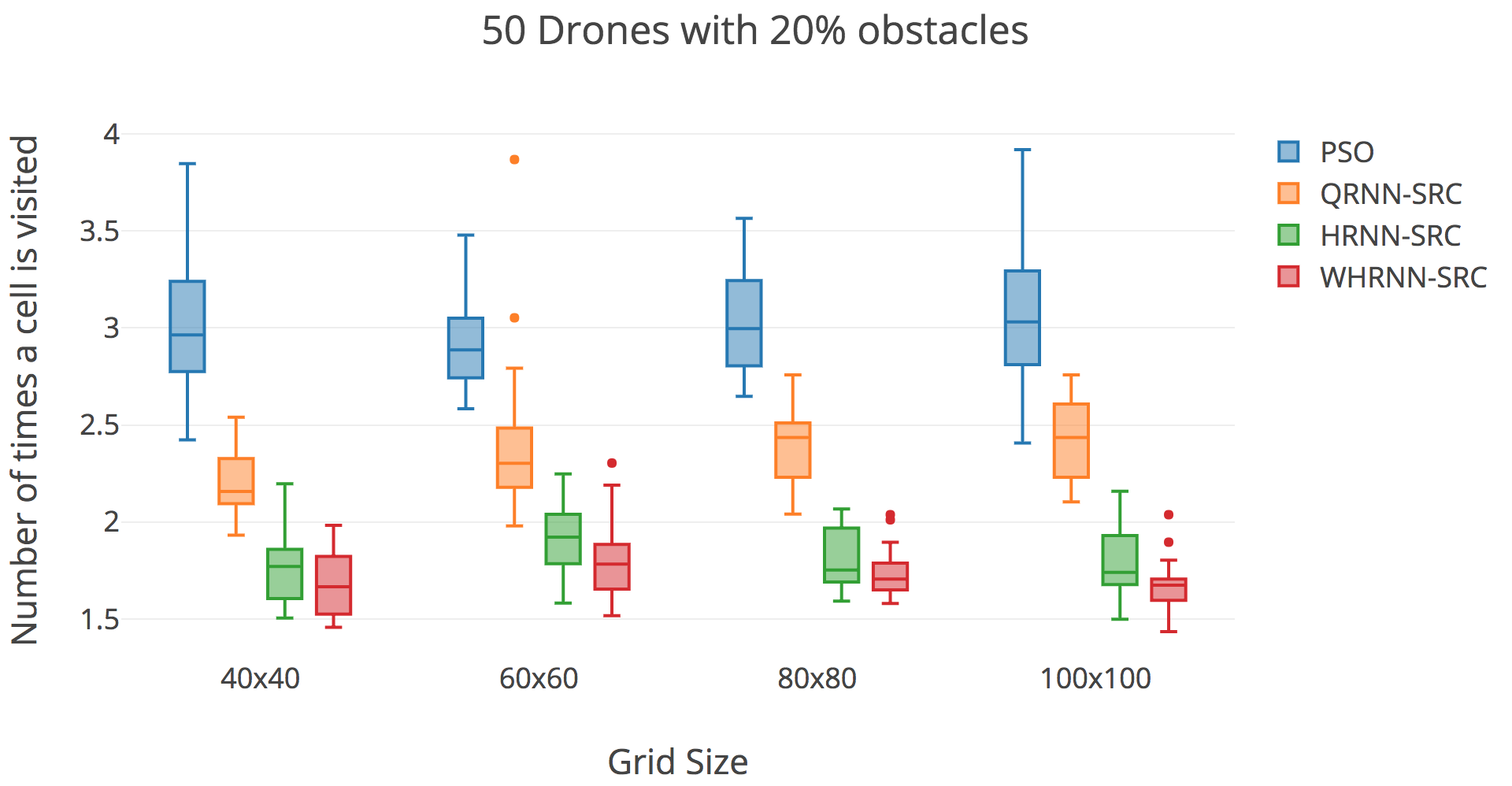}
\caption{Grid Size vs Average Re-Explored Cells.}\label{fig:Picture8}
}
\end{figure}

\autoref{fig:Picture8} represents the performance of Weighted RNN vs traditional PSO model, with an increase in grid size keeping the number of drones as constant. Here the approaches are compared taking 50 drones with an increase in the grid size. In this case also we can see that the weighted RNN outperforms the PSO based model. Our solution performs better not only in terms of the average number of times a cell is visited but also it has much less variance.

The above box plots are generated with extensive grid search on the parameters as shown in \autoref{tab:first}.
For each combination of number of drones and size of grid, the simulation environment is created 20 times with 20\% random obstacles. For each of the simulation environment grid search on parameters as shown in \autoref{tab:first} is performed.
For each combination of algorithm type, number of drones and size of grid the best parameters achieved will be different.
For instance, with an increase in the number of drones, the value of `$\alpha$' should increase to spread the drones farther away.

%
%
%

For all the experiments, parameters which give the best result after applying grid search are used.

\subsection{EVALUATION OF FAILURES}
In real time, it is very probable that failures might occur in the exploration process which can lead to massive delays in task completion.\par
There are two kinds of failures that can occur: \textbf{1. Temporary Failures:} Drones are affected by temporary failures due to weather and wind conditions, terrain features, communication latency, etc. The system can recover from such failures by incorporating a robust control algorithm into the drones \textbf{2. Permanent Failures in Drones: }Permanent failures in communication between drones can lead to massive task delays or permanent sensor failures can lead to partial exploration and mapping or motor and controller failures can lead to odometry/state estimation errors and accidents causing mid-air drone collisions.

\begin{figure}[htb]
{
\centering
\includegraphics[width=0.90\textwidth,height=0.40\textwidth]{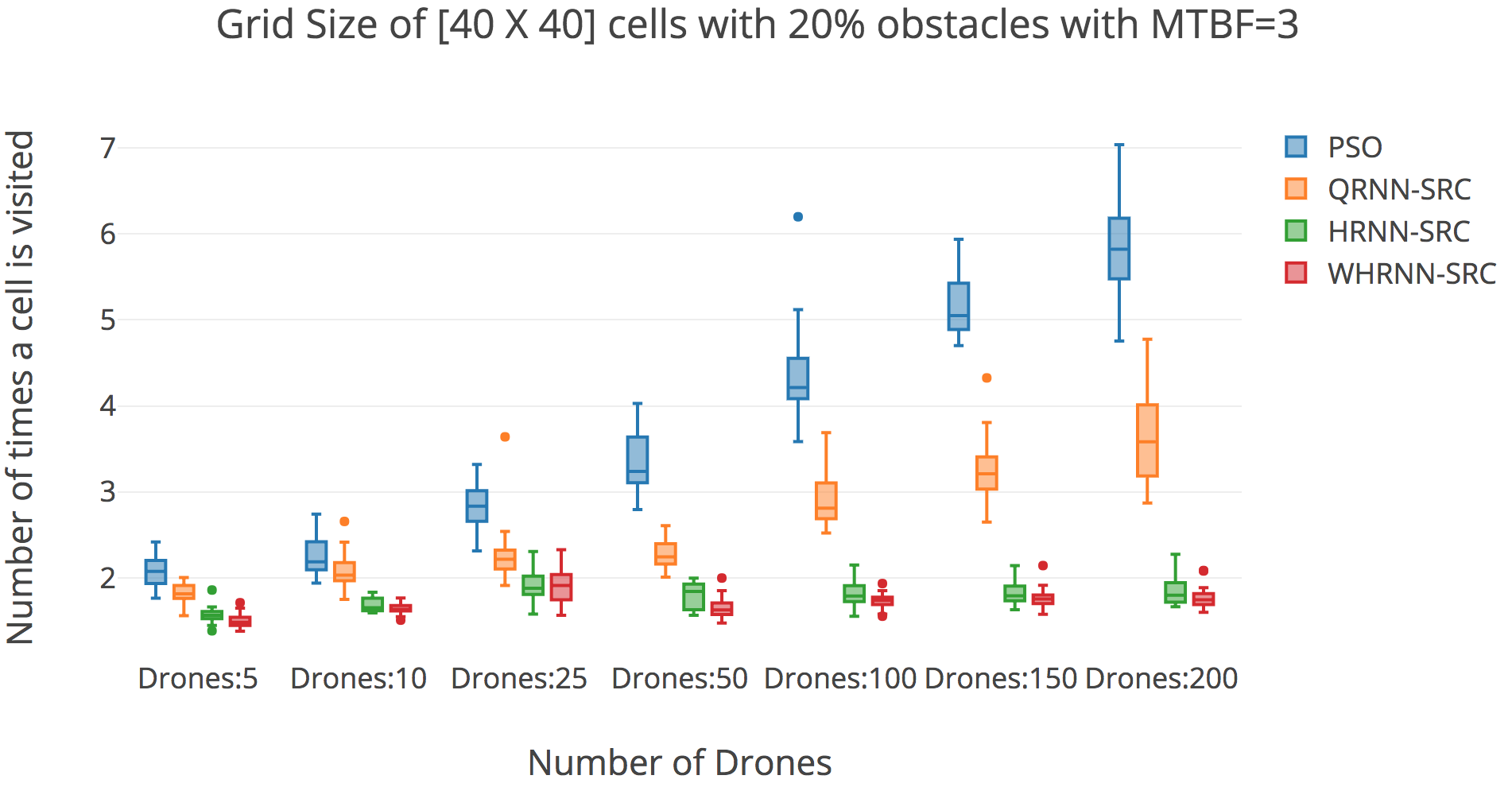}
\caption{Number of drones vs Avg Re-explored Cells with MTBF=3.}\label{fig:Picture9}
}
\end{figure}

Communication failures are usually simulated using MTBF (Mean Time Between Failures).
At the time \textit{`t'} the reliability of communication system R(t) for a drone is given by \eqref{eq:star5}:
\begin{equation}\label{eq:star5}
R(t) = e^{-\dfrac{t}{MTBF}}
\end{equation}
Failure in communication leads drones to re-exploration of the areas already explored by the companion drones. 
The plot as shown in \autoref{fig:Picture9} is simulated taking MTBF of 3 for all drones.

After a lapse in communication as soon as the communication starts working the drone gets the latest information regarding the area explored by the companion drones. The results above show that our algorithm outperforms the traditional PSO even in case of communication failures.  

\section{Summary}
In this paper, we improved upon the existing PSO algorithm for coordinating a swarm of drones while they are exploring an unknown environment. 
The goal of the paper is to reduce the redundancy and achieve complete map coverage.
Our proposed algorithm is scalable and outperforms even in case of failure in communication among drones.
Using two kinds of communication on a hexagonal grid with the weighted RNN algorithm improved the efficiency drastically. 
In the future, the same approach can also be applied to the continuous region rather than the discrete cells taken presently.

\nocite{*}
\bibliographystyle{scsproc}
\bibliography{demobib}

\section*{Author Biographies}

\textbf{\uppercase{AYUSH DATTA}} is a research student of "Agents and Applied Robotics Group" at "International Institute of Information Technology, Hyderabad". His interests lie in Robotics, Simulation and Path Planning. His email address is \email{ayush.datta@research.iiit.ac.in}.

\textbf{\uppercase{RAHUL TALLAMRAJU}} is a Ph.D. student of "Agents and Applied Robotics Group" at "International Institute of Information Technology, Hyderabad". His interests lie in Multi-Robot Systems, Multi-Agent Motion Planning. His email address is \email{rahul.t@research.iiit.ac.in}.

\textbf{\uppercase{KAMALAKAR KARLAPALEM}} is a Full Professor at "International Institute of Information Technology, Hyderabad". He is the head of Data Sciences and Analytics Centre and Agents and Applied Robotics Group. He holds a Ph.D. from (Georgia Institute of Technology, USA), His research interests include Database system, data visualization, data analytics, multi agent systems, workflows and electronic contracts. His email address is \email{kamal@iiit.ac.in}.
\end{document}